\documentclass[reqno]{amsart}
\usepackage{amsmath,amssymb}
\usepackage{xcolor}
\usepackage[mathletters]{ucs}
\usepackage[utf8x]{inputenc}
\usepackage{listings}
\usepackage{graphicx}
\usepackage{bm}
\usepackage[section]{placeins}

\newcommand{\comment}[1]{}
\newcommand{\perm}{\mbox{perm}}
\newcommand{\F}{\mathbb{F}}
\newcommand{\magn}{\mbox{mag}}
\newcommand{\sgn}{\mbox{sgn}}

\newtheorem{theorem}{Theorem}[section]
\newtheorem{conjecture}{Conjecture}[section]
\newtheorem{definition}{Definition}[section]

\begin{document}

\title{Fast computation of permanents over $\F_3$ via $\F_2$ arithmetic}
\author{Danny Scheinerman}
\address{Center for Communications Research, Princeton, NJ USA}
\email{daniel.scheinerman@gmail.com}

\begin{abstract} 
We present a method of representing an element of $\F_3^n$ as an element of $\F_2^n\times \F_2^n$ which in practice will be a pair of unsigned integers. We show how to do addition, subtraction and pointwise multiplication and division of such vectors quickly using primitive binary operations (and, or, xor). We use this machinery to develop a fast algorithm for computing the permanent of a matrix in $\F_3^{n\times n}$. We present Julia code for a natural implementation of the permanent and show that our improved implementation gives, roughly, a factor of 80 speedup for problems of practical size. Using this improved code, we perform Monte Carlo simulations that suggest that the distribution of $\perm(A)$ tends to the uniform distribution as $n\to\infty$.
\end{abstract}

\maketitle

\section{Introduction}

The permanent of a matrix $A$, denoted $\perm(A)$, is given by
\begin{equation}\label{eqn:perm}
\perm(A) = \sum_{\sigma \in S_n} \prod_{i=1}^n a_{i,\sigma(i)}.
\end{equation}
This is reminiscent of the formula for the determinant, except we do not weight the terms in the sum with the sign of the permutation. The determinant can be computed in polynomial time via, for example, Gaussian elimination. Computing the permanent, however, seems much harder. It is known~\cite{Valiant} that computing the permanent of a $0,1$-matrix is $\#\mbox{P}$-complete. In characteristic $2$, where the permanent and determinant coincide, the computation is polynomial time, but for other moduli it appears hard in general. In \cite{Kogan} it is shown that if $A\in \F_3^{n\times n}$ is such that $\mbox{rank}(AA^T-I)\le 1$ then the permanent can be computed in polynomial time, but otherwise is $\#\mbox{P}(\mbox{mod}~3)$-complete. In general, one can employ Equation \eqref{eqn:perm} to give an algorithm that takes roughly $n!$ operations to compute the permanent, but Ryser~\cite{Ryser} found the significant improvement:
\begin{equation}\label{eqn:ryser}
\perm(A) = (-1)^n \sum_{S\subseteq\{1,\ldots,n\}} (-1)^{|S|} \prod_{i=1}^n \sum_{j\in S} a_{i,j}.
\end{equation}
In words, Equation~\eqref{eqn:ryser} says that if you think of $A$ as $n$ row vectors, and take all $2^n$ subset sums of these vectors and for each subset sum take the product of the entries, then summing these products with the appropriate signs gives you the permanent of your matrix. Ryser's formula then naturally leads to an algorithm using $O(n^2 2^n)$ field operations. This can be improved by traversing the subsets of $\{1,\ldots,n\}$ in Gray code order thus requiring a single row addition or subtraction at each iteration. This improves the running time to $O(n 2^n)$.

In this short note we are interested in quickly computing the permanent of a matrix over $\F_3$. We will present an approach that has the same asymptotic running time as above, but by using $\F_2$ arithmetic which computer architectures are well suited to, gives a considerable speedup. To make the comparison we will present a natural implementation (see \verb+permanent_Ryser+ in Section~\ref{sec:comp}) of the Gray coded version of Ryser's formula in the programming language Julia.\footnote{We provide Julia code both so that an interested reader can easily experiment with these computations, but also the syntax of Julia is human friendly and so serves well as pseudocode.} For a $36\times 36$ matrix this code took $8857.9$ seconds (about 2.5 hours) to compute the permanent on a 4.20GHz desktop. Using the machinery we will develop, we implement \verb+permanent_mod3+ (also found in Section~\ref{sec:comp}) which required 101.9 seconds to compute the same permanent. This is a factor of $86.9$ speedup. It is straightforward to parallelize Ryser's formula. We extrapolate that it would take about 64 thousand hours (about 7.3 cpu years) to compute the permanent of a $50\times 50$ matrix using \verb+permanent_Ryser+ whereas we were able to compute the permanent of a $50\times 50$ matrix over $\F_3$ using \verb+permanent_mod3+ in  832 hours of parallel compute time.

This project was motivated by considering the following question. Does the distribution of the permanent of a random matrix in $\F_3^{n\times n}$ approach the uniform distribution as $n\to \infty$? A natural first step is to collect data via Monte Carlo simulations. The machinery we develop greatly speeds up these simulations. Results of such simulations can be found in Section~\ref{sec:distribution}. These results do strongly suggest an affirmative answer to the above question, reflected in Conjecture~\ref{conj:uniform}.

\section{Representing elements of $\F_3^n$ as elements in $\F_2^n\times \F_2^n$}\label{sec:reps}

Our main idea is to assign to the elements of $\F_3$ a representation as a pair of values in $\F_2$. So we define a map $\phi:\F_3 \to \F_2\times \F_2$. We want to do this in a way such that if $\alpha,\beta \in \F_3$, then given the $\F_2\times\F_2$ representations of $\alpha$ and $\beta$, we can compute the representations of $\alpha+\beta$, $\alpha-\beta$, $\alpha \beta$ and $\alpha/\beta$ in as few binary operations (and, or, xor) as possible. Note that to compute the permanent we will not need multiplication or division, but we include them for completeness. Moreover, if $\bm{\alpha} = (\alpha_1,\ldots,\alpha_n)\in \F_3^n$ we will represent $\bm{\alpha}$ as an element of $\F_2^n \times \F_2^n$. Provided $n\le 64$, this architecturally will be a pair of unsigned integers. Then armed with these representations of elements $\bm{\alpha}, \bm{\beta} \in \F_3^n$ we can, for example, find the representation of $\bm{\alpha} + \bm{\beta}$ by using primitive binary operations where our ``atoms'' are unsigned 64-bit integers. 

Formally, we define $\phi:\F_3 \to \F_2\times\F_2$ via
\begin{equation}
\phi(0) = (0,0),\hphantom{abc}\phi(1)=(1,0),\hphantom{abc}\phi(-1)=(1,1).
\end{equation}
We think of $\phi$ as mapping an element in $\F_3$ to a pair $(\magn,\sgn)$ where $\magn$ is the ``magnitude'' of the element (0 if it is zero and 1 if it is nonzero) and the ``sign'' differentiates between $1$ and $-1$. Although not in the image of $\phi$, we will think of $(0,1)\in \F_2\times\F_2$ as an alternative representation of $0\in\F_3$ and thus define the ``inverse'' map $\psi:  \F_2\times \F_2 \to \F_3$ by
\begin{equation}
\psi((0,0))=\psi((0,1))=0,\hphantom{abc}\psi((1,0))=1,\hphantom{abc}\psi((1,1))=-1.
\end{equation}
Clearly, we have for any $\alpha\in\F_3$ that $\psi(\phi(\alpha))=\alpha$.

As stated above we extend $\phi$ pointwise so as to map elements of $\F_3^n$ to elements of $\F_2^n \times \F_2^n$. So for example if $\bm{\alpha} = (1,1,0,-1)\in \F_3^4$, then $\phi(\bm{\alpha}) = \big((1,1,0,1), (0,0,0,1)\big).$
On a computer, we then may represent $\bm \alpha$ as the pair of unsigned integers \texttt{(0xd,0x1)}. However, as stated we consider $(0,1)$ a valid reprentation of $0$ and thus have \texttt{(0xd,0x3)} as an alternative representation.
 \begin{definition}
 For $\bm \alpha\in\F_3^n$ we say a pair of vectors $(\magn,\sgn)\in\F_2^n$ is a \emph{bipedal representation} of $\bm \alpha$ provided $\psi((\magn,\sgn)) = \bm{\alpha}$.
 \end{definition}
 Thus for any $\bm \alpha\in\F_3^n$, $\phi(\bm \alpha)$ is one (of possibly several) bipedal representation of $\bm \alpha$.

Given bipedal representations $(\magn_1,\sgn_1)$ and $(\magn_2,\sgn_2)$ for $\bm \alpha$ and $\bm \beta$ respectively, one naturally wants formulas for bipedal representations of $\bm{\alpha}+\bm{\beta}$, $\bm{\alpha}-\bm{\beta}$, $\bm{\alpha}\times\bm{\beta}$ and $\bm{\alpha}/\bm{\beta}$. Here $\bm{\alpha}\times\bm{\beta}$ and $\bm{\alpha}/\bm{\beta}$ are the pointwise product and ratio\footnote{For division we will only be concerned that our formulas give sensible answers when the denominator is nonzero.} respectively. If we consider addition,  then Table~\ref{tab:addtruth} gives the partial truth table that $\magn_+$ and $\sgn_+$ must satisfy. It is a partial truth table as we are indifferent as to which representation of $0\in\F_3$ is returned. The values displayed in red are those given by the formula in Theorem~\ref{thm:ops}. Similar flexibility holds for subtraction, multiplication and division. We will exploit this flexibility to seek the truth table for which we are able to find the most efficient implementations.

\begin{table}[ht!]
\caption{Truth table for addition. Values in red are those given by the formula in Theorem~\ref{thm:ops}.}
\begin{center}
\begin{tabular}{|c||c|c|c|c||c|c|}
\hline
$\F_3$ equation & $\magn_1$ &  $\sgn_1$ &  $\magn_2$ &  $\sgn_2$ &  $\magn_+$ &  $\sgn_+$ \\
\hline
$0+0=0$  &  0   &   0   &   0   &   0   &   0   &   \color{red}{0}       \\
$0+0=0$  &  0   &   0   &   0   &   1   &   0   &   \color{red}{0}        \\
$0+0=0$  &  0   &   1   &   0   &   0   &   0   &   \color{red}{1}         \\
$0+0=0$  &  0   &   1   &   0   &   1   &   0   &   \color{red}{1}         \\
$0+1=1$  &  0   &   0   &   1   &   0   &   1   &   0       \\
$0+1=1$  &  0   &   1   &   1   &   0   &   1   &   0       \\
$0+2=2$  &  0   &   0   &   1   &   1   &   1   &   1       \\
$0+2=2$  &  0   &   1   &   1   &   1   &   1   &   1       \\
$1+1=2$  &  1   &   0   &   1   &   0   &   1   &   1       \\
$1+0=1$  &  1   &   0   &   0   &   0   &   1   &   0       \\
$1+0=1$  &  1   &   0   &   0   &   1   &   1   &   0       \\
$1+2=0$  &  1   &   0   &   1   &   1   &   0   &   \color{red}{0}         \\
$2+0=2$  &  1   &   1   &   0   &   0   &   1   &   1       \\
$2+0=2$  &  1   &   1   &   0   &   1   &   1   &   1       \\
$2+1=0$  &  1   &   1   &   1   &   0   &   0   &   \color{red}{1}         \\
$2+2=1$  &  1   &   1   &   1   &   1   &   1   &   0       \\
\hline
\end{tabular}
\end{center}
\label{tab:addtruth}
\end{table}

For addition, for example, one can exhaust over the $2^6=64$ choices for the representation of zero and use a computer program to seek a circuit that computes the function in a minimal number of operations (and, or, xor). For addition and subtraction there were several choices that had equally efficient circuits (6 operations). In these cases we choice one arbitrarily. These formulas are given in Theorem~\ref{thm:ops}. We will use $\&$ to denote the ``and'' operation, $|$ to denote the ``or'' operation and $\oplus$ to denote\footnote{Note that in Julia the xor operation is denoted by the unicode character \texttt{⊻}. This will be reflected in the subsequent provided code.}  ``xor.''

\begin{theorem}\label{thm:ops}
Let $\bm \alpha$ and $\bm \beta$ have bipedal representations $(\magn_1,\sgn_1)$ and $(\magn_2,\sgn_2)$ respectively. Then bipedal representations $(\magn_+,\sgn_+)$, $(\magn_-,\sgn_-)$, $(\magn_\times,\sgn_\times)$ and $(\magn_\div,\sgn_\div)$ for $\bm{\alpha}+\bm{\beta}$, $\bm{\alpha}-\bm{\beta}$, $\bm{\alpha}\times \bm{\beta}$ and $\bm{\alpha}/\bm{\beta}$ respectively are given by
\begin{align*}
\magn_+ &= (\magn_2~\&~(\magn_1\oplus \sgn_1\oplus\sgn_2)) ~|~ (\magn_1 \oplus \magn_2)  \\
\sgn_+ &= (\magn_2~\&~(\magn_1\oplus \sgn_1\oplus\sgn_2)) \oplus \sgn_1 \\
\magn_- &= (\magn_1~\&~(\sgn_1 \oplus \sgn_2)) ~|~ (\magn_1 \oplus \magn_2)  \\ 
\sgn_- &= (\magn_1~\&~(\sgn_1 \oplus \sgn_2)) \oplus (\magn_2 \oplus \sgn_2) \\
\magn_\times &= \magn_1 ~\&~\magn_2 \\
\sgn_\times &= \sgn_1 \oplus \sgn_2 \\
\magn_\div &= \magn_1 \\
\sgn_\div &= \sgn_1 \oplus \sgn_2.
\end{align*}
\end{theorem}

At a glance it seems the formula for addition requires $9$ Boolean operations and subtraction requires $8$. However, there is a common term in the formulas for $\magn_+$ and $\sgn_+$, namely $\magn_2~\&~(\magn_1\oplus \sgn_1\oplus\sgn_2)$. If we compute this once (using 3 operations) and remember it then we only need 3 additional operations to compute the above formulas. So, as implemented, addition uses $6$ operations. Similarly, subtraction has the common term $\magn_1~\&~(\sgn_1 \oplus \sgn_2)$. So subtraction can be accomplished in $6$ operations as well. Multiplication is just two operations. For division, we have only need one operation. Implementations of these operations in Julia can be found in Appendix~\ref{sec:code}.

Theorem~\ref{thm:ops} can of course be proved simply by checking the $16$ possible input values for each operation. Some of these formulas may look strange as, for example, one might expect addition to be commutative and therefore the formulas more symmetric, but as we are agnostic to the choice of which representation for zero to output this need not be the case.

\section{ Computing the permanent}\label{sec:comp}

To make a point of comparison, here is (we think) a reasonable implementation of a Gray coded\footnote{We use the Reflected Binary Gray code (RBC) in our implementation. For this code, in the $i$-th step we flip bit number $\mbox{tz}(i)$ where $\mbox{tz}$ counts the number of trailing zeros.} version of Ryser's formula in Julia\footnote{Note that Julia employs $1$-based indexing. Thus the ``\texttt{t+1}'' on lines 12 and 14 of \texttt{permanent\_Ryser}.}:

\begin{verbatim}
 1|  #compute permanent of integer matrix
 2|  function permanent_Ryser(A::Array{Int64,2})
 3|      n = size(A,1) #assume A is square
 4|      s = 0 #to accumulate the products of the row subset sums
 5|      x = UInt64(0) #to maintain the state of the Gray code
 6|      v = zeros(Int64,n) #the current subset sum
 7|      for i in 1:2^n-1
 8|          t = trailing_zeros(i) #next bit to flip in Gray code
 9|          b = (x>>t)&1 #if 1 we do addition else subtraction
10|          x = xor(x,1<<t) #update Gray code state
11|          if b == 1
12|              v += A[t+1,:] #add (t+1)-st row to v
13|          else
14|              v -= A[t+1,:] #subtract (t+1)-st row from v
15|          end
16|          s += (-1) ^ (i&1) * prod(v)
17|      end
18|      s *= (-1)^n
19|      return s
20|  end
\end{verbatim}

For our application we are only concerned with computing the permanent modulo $3$. For modest sizes if one inputs the entries of $A$ as $-1$, $0$, or $1$ then subset row sums tend to have small entries and so integer overflow is not a concern. However, if we always reduce the vector specified by the variable ``\texttt{v}'' to lie in $\{-1,0,1\}^n$ then its product lies in $\{-1,0,1\}$ and so overflow is never an issue. So we define the centered mod operation:
\begin{verbatim}
 1|  function cmod3(x::Int64)
 2|      a = mod(x,3)
 3|      return (a==2 ? -1 : a)
 4|  end
\end{verbatim}
and in \verb+permanent_Ryser+ we can replace line 16 with:
\begin{verbatim}
16|          s += (-1) ^ (i&1) * prod(cmod3.(v))
\end{verbatim}
which, experimentally, has no discernible effect on the run time.

In a test run on a 4.2 GHz machine we ran \verb+permanent_Ryser+ on random matrices with elements in $\{-1,0,1\}$. The average run times are given in Table~\ref{tab:ryser}. The run times, $T(n)$, are well approximated by the estimate
\begin{equation}\label{eqn:est_old}
T(n) \approx 4.1\times 10^{-9} \times n \times 2^n~\text{seconds}.
\end{equation}

\begin{table}[ht!]
\caption{Average run time (in seconds) for \texttt{permanent\_Ryser} and \texttt{permanent\_mod3} for various values of $n$.}
  \begin{center}
  \begin{tabular}{|c|c|c|}
\hline
$n$ & \verb+permanent_Ryser+ & \verb+permanent_mod3+\\
\hline
24 & 1.96  &  0.025  \\
26 & 7.98  &  0.099  \\
28 & 32.3  &  0.401  \\
30 & 131.5 &  1.59  \\
32 & 533.5 &  6.34  \\
34 & 2166.9 & 25.52 \\
36 & 8857.9 & 101.9 \\
\hline
  \end{tabular}
  \end{center}
  \label{tab:ryser}
\end{table}

We now will use the techniques developed in Section~\ref{sec:reps} to greatly speed up our computation of the permanent mod 3. Given an $n\times n$ matrix, $A$, where $n\le 64$ we can apply our map $\phi$ to the rows so as to represent the matrix as a list of $n$ pairs of unsigned 64-bit integers. We initialize the bipedal representation of the zero vector (a pair of $\F_2$ zero vectors) and proceed to Gray code through the subsets of the rows adding or subtracting a row as appropriate. These computations are all done using the formulas given in Theorem~\ref{thm:ops}. At each iteration, Ryser's formula tells us we need to take the product of the subset row sum. Since this is a product of $n$ elements in $\F_3$ there is a good chance the product is zero. Heuristically, there is only a $(2/3)^n$ chance that it is nonzero.\footnote{Exploiting this fact to speed up the computation is a quite interesting question that we have investigated separately, but not the subject of the present note.} This can be checked very quickly by looking at the magnitude value for the subset row sum. If the relevant bits (the $n$ least significant bits) of the $\magn$ value are not identically equal to one, then it contains a zero and the product is zero and we thus need not update our running sum. Otherwise, the product is $-1$ raised to the number of $-1$'s which is given by the number of $1$'s in the $\sgn$ value\footnote{We use the Julia function \texttt{count\_ones} to compute the popcount. All we actually need is the poppar, the parity of the popcount. Since this code is relatively rarely called we do not worry about implementing this potential improvement.}. This is implemented in the function \verb+permanent_mod3+ found below. The companion functions that implement $\phi$, $\psi$, addition, subtraction and multiplication can be found in the appendix.

In a test on the same 4.2 GHz machine as before we ran \verb+permanent_mod3+ for random $\{-1,0,1\}$ matrices. The timings are found in 
Table~\ref{tab:ryser}. Since the operations at each step are word operations they are fast and constant time. Thus this implementation is effectively\footnote{Of course it is still $O(n 2^n)$ asymptotically. The number of 64-bit integers needed to represent the rows is $\lceil n/64\rceil$ but for feasible values of $n$ this equals 1 or, ambitiously, 2.} $O(2^n)$ rather than $O(n 2^n)$. The run times for \verb+permanent_mod3+, denoted by $T'(n)$, are closely approximated by the estimate
\begin{equation}\label{eqn:est_new}
T'(n) \approx 1.5\times 10^{-9} \times 2^n~\text{seconds}.
\end{equation}
Of course if one were to extend the data far enough a discontinuity would be seen as we move from $n=64$ to $n=65$ and this estimate would break down.

\begin{verbatim}
 1|  #compute the permanent mod 3 of integer matrix, A
 2|  #assumes entries of A are -1, 0 or 1
 3|  function permanent_mod3(A::Array{Int64,2})
 4|      n = size(A,1) #assume A is square
 5|      @assert n <= 64 #so mag, sgn fit in single UInt64
 6|      ROWS = [F3_vector_to_bipedal_rep(A[i,:]) for i in 1:n]
 7|      mask = (UInt64(1) << n) - 1
 8|      s = 0
 9|      x = UInt64(0)
10|      mag, sgn = F3_vector_to_bipedal_rep(zeros(Int64,n))
11|      for i in 1:2^n-1
12|          t = trailing_zeros(i) #next bit to flip in Gray code
13|          b = (x>>t)&1 #if 1 we do addition else subtraction
14|          x = xor(x,1<<t) #update Gray code state
15|          if b == 1
16|              #subtract (t+1)-st row
17|              mag, sgn = sub_reps(mag,sgn,ROWS[t+1]...)
18|          else
19|              #add (t+1)-st row
20|              mag, sgn = add_reps(mag,sgn,ROWS[t+1]...)
21|          end
22|          if mag == mask  #so no zeros in subset row sum
23|              s += (-1) ^ (count_ones(sgn) + (i&1))
24|          end
25|      end
26|      s *= (-1)^n
27|      return cmod3(s)
28|  end    
\end{verbatim}

\subsection{Parallel computation: a large example}

The above computations are straightforward to run in parallel. Each job will accumulate the sum coming from a block of consecutive Gray code states. One just needs to initialize properly by jumping\footnote{For the RBC, state $i$ is given by \texttt{i⊻(i>>1)}.} to the appropriate state to initialize the row subset sum. As a demonstration of the size of problem this makes feasible, let $\Pi_n$ be the $n\times n$ matrix whose $i,j$ entry is the $(n-1)i+j$-th digit of $\pi$. So the first row are the first $n$ digits of $\pi$, the second row has digits $n+1,\ldots,2n$ and so on. To compute $\perm(\Pi_{50})$ modulo $3$ using \verb+permanent_Ryser+ we estimate using Equation~\eqref{eqn:est_old} approximately  64 thousand hours (7.3 years) of compute time (this could of course be done in parallel). However, a parallel implementation of \verb+permanent_mod3+ determined that $\perm(\Pi_{50})\equiv -1 \pmod 3$ in just 832 hours of total computation.

\section{Distribution of permanent of random matrix over $\F_3$}\label{sec:distribution}

In this section we give the results of numerical experiments that suggests that the distribution of the permanent of a random matrix in $\F_3^{n\times n}$ approaches the uniform distribution as $n\to \infty$. For small values of $n$ we can exactly compute the distribution and for larger values of $n$ we do Monte Carlo simulations. Notice for any $n$ we have a bijection between the matrices with permanent equal to $1$ and those with permanent equal to $-1$ given by negating the first row. Thus for any $n$ the total counts among the $3^{n^2}$ matrices in $\F_2^{n\times n}$ with permanent equal to $1$ versus $-1$ are the same. So it suffices to find/estimate the counts (or probabilities) of having the permanent equal $0$. For any positive integer $n$ we let $z(n)=|\{A\in \F_3^{n\times n} : \perm(A) = 0\}|$. For $n\le 5$ we compute $z(n)$ exactly. This is given in table~\ref{tab:smalln}.

\begin{table}[ht!]
\caption{Number of matrices $A\in \F_3^{n\times n}$ such that $\perm(A)=0$ for small $n$.}
  \begin{center}
  \begin{tabular}{|c|l|c|}
\hline
$n$ & $z(n)$ & $ \vphantom{a^{b^{c^d}}} z(n)/3^{n^2}$ \\
\hline
1  & 1      & 0.3333 \\
2  & 33      & 0.4074  \\
3  & 8163    &   0.4147   \\
4  & 17116353  &  0.3976  \\ 
5  & 317193401763 &  0.3744   \\
\hline
  \end{tabular}
  \end{center}
  \label{tab:smalln}
\end{table}

\begin{figure}[!htb]\label{fig:probzero}
\caption{Probability a random matrix in $\F_3^{n\times n}$ has permanent equal to zero. Points in blue are exact, points in red are estimated from Monte Carlo trials.}
\begin{center}
\scalebox{0.8}{\includegraphics{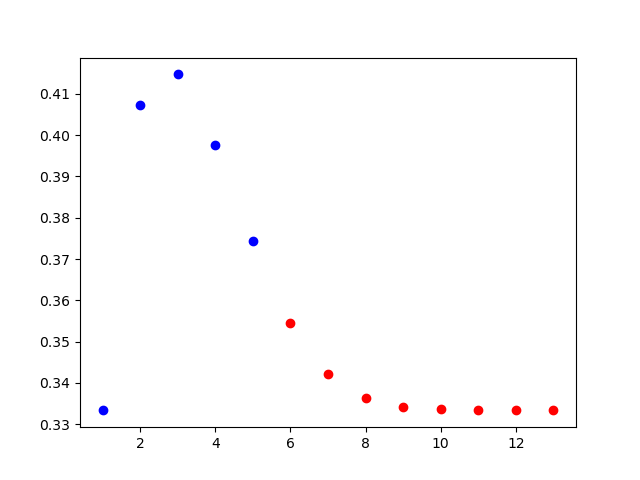}} 
\end{center}
\end{figure}

For larger values of $n$ we perform Monte Carlo simulations to estimate the proportion of matrices with permanent equal to zero. These are reflected in table~\ref{tab:largen}. For $n\le 13$ the proportions are statistically distinguishable from $1/3$, but for larger $n$ they are not. A plot of these probabilities for $n\le 13$ is found in Figure~\ref{fig:probzero}. These results motivate Conjecture~\ref{conj:uniform}. These simulations took about one full day on $128$ processors.

\begin{conjecture}\label{conj:uniform}
As $n\to\infty$ the distribution of $\perm(A)$ for $A\in \F_3^{n\times n}$ approaches the uniform distribution. Equivalently,
\begin{equation}
\lim_{n\to\infty} \frac{z(n)}{3^{n^2}} = \frac 13.
\end{equation}
\end{conjecture}

\begin{table}[h!]
\caption{ Results of Monte Carlo experiments for the probability a randomly chosen matrix in $\F_3^{n\times n}$ has permanent equal to 0.}\label{tab:largen}
\begin{center}
\begin{tabular}{|c|l|c|}
\hline
$n$ & zero count & $\log_{10}(\#~\mbox{trials})$ \\
\hline
6       &       35456365448     &       11      \\
7       &       34209345718     &       11      \\
8       &       33623043873     &       11      \\
9       &       33417515901     &       11      \\
10      &       33358878343     &       11      \\
11      &       3334206857      &       10      \\
12      &       3333537904      &       10      \\
13      &       3333483177      &       10      \\
14      &       3333394825      &       10      \\
15      &       333332350       &       9       \\
16      &       333308622       &       9       \\
17      &       333314098       &       9       \\
18      &       33331991        &       8       \\
19      &       33338438        &       8       \\
20      &       33338902        &       8       \\
21      &       3332782 &       7       \\
22      &       3333672 &       7       \\
23      &       3336968 &       7       \\
24      &       3333518 &       7       \\
25      &       3332961 &       7       \\
26      &       3335524 &       7       \\
27      &       3332955 &       7       \\
28      &       333743  &       6       \\
29      &       334097  &       6       \\
30      &       333080  &       6       \\
\hline
\end{tabular}
\end{center}
\end{table}

\section{Acknowledgments}

Thanks to Steven Fischer, Keith Frankston and Miller Maley for helpful discussions. Steven Boyack and Marshall Buck read early versions of this paper and gave helpful comments.

\appendix

\section{Julia Code}\label{sec:code}

In this appendix we give Julia code that implements the maps $\phi$ and $\psi$ as well as addition, subtraction, multiplication and division of vectors in bipedal form.

The following is an implementation of the map $\phi:\F_3 \to \F_2\times \F_2$ described in Section~\ref{sec:reps}. It takes an array of $\{-1,0,1\}$ values and returns a pair of unsigned 64-bit integers. 
\begin{verbatim}
 1|  #Convert w, a vector of {-1,0,1}'s, to a pair of UInt64's
 2|  function F3_vector_to_bipedal_rep(w::Array{Int64,1})
 3|      @assert (minimum(w)>=-1) && (maximum(w)<=1)
 4|      n = length(w)
 5|      @assert n <= 64
 6|      mag = UInt64(0)
 7|      sgn = UInt64(0)
 8|      for i in 1:n
 9|          if w[i] != 0
10|              mag += (1<<(n-i))
11|              if w[i] == -1
12|                  sgn += (1<<(n-i))
13|              end
14|          end
15|      end
16|      return mag, sgn
17|  end
\end{verbatim}

The following is an implementation of the map $\psi: \F^n_2\times \F^n_2 \to \F^n_3$  described in Section~\ref{sec:reps}. It takes a pair of unsigned 64-bit integers $\magn$ and $\sgn$ and an integer $n$ and returns the corresponding length $n$ array, $w$, with elements in $\{-1,0,1\}$. This function is unused in our implementation of \verb+permanent_mod3+ but we include it for reference.
\begin{verbatim}
 1|  #given UInt64's mag and sgn and Int64 n specifying length
 2|  #returns the corresponding {-1,0,1}-vector
 3|  function bipedal_rep_to_F3_vector(mag::UInt64,sgn::UInt64,n::Int64)
 4|      w = zeros(Int64,n)
 5|      for i in 1:n
 6|          if (mag >> (n-i)) & 1 == 1
 7|              w[i] = (-1) ^ ((sgn>>(n-i)) & 1)
 8|          end
 9|      end
10|      return w
11|  end
\end{verbatim}

The following is an implementation of addition based on the formulas given in Theorem~\ref{thm:ops}.
\begin{verbatim}
 1|  #addition of two bipedal representations
 2|  function add_reps(mag1::UInt64,sgn1::UInt64,mag2::UInt64,sgn2::UInt64)
 3|      x = mag2 & (mag1 ⊻ sgn1 ⊻ sgn2)
 4|      mag3 = x | (mag1 ⊻ mag2)
 5|      sgn3 = x ⊻ sgn1
 6|      return mag3, sgn3
 7|  end
\end{verbatim}

The following is an implementation of subtraction based on the formulas given in Theorem~\ref{thm:ops}.
\begin{verbatim}
 1|  #subtraction of two bipedal representations
 2|  function sub_reps(mag1::UInt64,sgn1::UInt64,mag2::UInt64,sgn2::UInt64)
 3|      x = mag1 & (sgn1 ⊻ sgn2)
 4|      mag3 = x | (mag1 ⊻ mag2)
 5|      sgn3 = x ⊻ (mag2 ⊻ sgn2)
 6|      return mag3, sgn3
 7|  end
\end{verbatim}

The following is an implementation of multiplication based on the formulas given in Theorem~\ref{thm:ops}.
\begin{verbatim}
 1|  #multiplication of two bipedal representations
 2|  function mul_reps(mag1::UInt64,sgn1::UInt64,mag2::UInt64,sgn2::UInt64)
 3|      mag3 = mag1 & mag2
 4|      sgn3 = sgn1 ⊻ sgn2
 5|      return mag3, sgn3
 6|  end
\end{verbatim}

The following is an implementation of division based on the formulas given in Theorem~\ref{thm:ops}.
\begin{verbatim}
 1|  #division of two bipedal representations
 2|  function div_reps(mag1::UInt64,sgn1::UInt64,mag2::UInt64,sgn2::UInt64)
 3|      mag3 = mag1
 4|      sgn3 = sgn1 ⊻ sgn2
 5|      return mag3, sgn3
 6|  end
\end{verbatim}

\bibliographystyle{plain}
\bibliography{F3perm.bib}

\begin{thebibliography}{1}

\bibitem{Kogan}
Grigory Kogan.
\newblock Computing permanents over fields of characteristic {$3$}: where and
  why it becomes difficult (extended abstract).
\newblock In {\em 37th {A}nnual {S}ymposium on {F}oundations of {C}omputer
  {S}cience ({B}urlington, {VT}, 1996)}, pages 108--114. IEEE Comput. Soc.
  Press, Los Alamitos, CA, 1996.

\bibitem{Ryser}
Herbert~John Ryser.
\newblock {\em Combinatorial mathematics}.
\newblock The Carus Mathematical Monographs, No. 14. Mathematical Association
  of America; distributed by John Wiley and Sons, Inc., New York, 1963.

\bibitem{Valiant}
L.~G. Valiant.
\newblock The complexity of computing the permanent.
\newblock {\em Theoret. Comput. Sci.}, 8(2):189--201, 1979.

\end{thebibliography}

\end{document}